\documentclass[preprint,12pt]{elsarticle}

\usepackage{amssymb}
\usepackage{amsmath}
\usepackage{multirow}
\usepackage{ulem}
\usepackage{threeparttable}
\usepackage{hyperref}
\usepackage{threeparttable}  
\usepackage[figuresright]{rotating} 
\usepackage{color}
\usepackage{microtype}

\journal{Journal of Systems and Software}

\begin{document}

\begin{frontmatter}



\title{MPLinker: Multi-template Prompt-tuning with Adversarial Training for Issue-commit Link Recovery}

\author[label1,label2]{Bangchao Wang}

\author[label1]{Yang Deng}
\author[label1,label2]{Ruiqi Luo}
\author[label3,label4]{Peng Liang}
\author[label5]{Tingting Bi}
\affiliation[label1]{organization={School of Computer Science and Artificial Intelligence},
            addressline={Wuhan Textile University},
            city={Wuhan},
            country={China}}

\affiliation[label2]{organization={Engineering Research Center of Hubei Province for Clothing Information},
            addressline={Wuhan Textile University},
            city={Wuhan},
            country={China}}

\affiliation[label3]{organization={School of Computer Science},
            addressline={Wuhan University},
            city={Wuhan},
            country={China}}

\affiliation[label4]{organization={Hubei Luojia Laboratory},
            city={Wuhan},
            country={China}}
  \affiliation[label5]{organization={The University of Western Australia},
            country={Australia}}

\begin{abstract}
In recent years, the pre-training, prompting and prediction paradigm, known as prompt-tuning, has achieved significant success in Natural Language Processing (NLP). Issue-commit Link Recovery (ILR) in Software Traceability (ST) plays an important role in improving the reliability, quality, and security of software systems. The current ILR methods convert the ILR into a classification task using pre-trained language models (PLMs) and dedicated neural networks. these methods do not fully utilize the semantic information embedded in PLMs, resulting in not achieving acceptable performance. To address this limitation, we introduce a novel paradigm: \textbf{Multi-template Prompt-tuning} with adversarial training for issue-commit \textbf{Link} recovery (MPLinker). MPLinker redefines the ILR task as a cloze task via template-based prompt-tuning and incorporates adversarial training to enhance model generalization and reduce overfitting. We evaluated MPLinker on six open-source projects using a comprehensive set of performance metrics. The experiment results demonstrate that MPLinker achieves an average \textcolor{black}{F1-score} of 96.10\%, Precision of 96.49\%, Recall of 95.92\%, MCC of 94.04\%, AUC of 96.05\%, and ACC of 98.15\%, significantly outperforming existing state-of-the-art methods. Overall, MPLinker improves the performance and generalization of ILR models, and introduces innovative concepts and methods for ILR. The replication package for MPLinker is available at \url{https://github.com/WTU-intelligent-software-development/MPLinker}.
\end{abstract}




\end{frontmatter}


\section{Introduction}
\label{Introduction}

Software Traceability (ST) activities link different software artifacts produced in software development to enhance developers' observability in practice, which supports diverse software engineering activities such as impact analysis, security assurance, and safety validation of software systems~\cite{lin2021traceability}. In particular, trace links between issues and commits (Issue-commit Link Recovery (ILR)) play a key role in bug localization~\cite{rahman2013sample}, defect prediction~\cite{kim2007predicting}, and other software maintenance tasks. In software development, developers commonly establish links between issues and commits by recording the relevant commit after resolving an issue. However, this process is not mandatory, relying on developers' voluntary actions, and manual link creation is time-intensive and demanding. As a result, developer negligence or other factors during the development process may lead to numerous missing links, impacting the overall quality and reliability of software systems.

To alleviate the above problems, researchers have increasingly focused on studying automated ILR methods, particularly learning-based ILR methods~\cite{nguyen2012multi, sun2017frlink, sun2017improving, mazrae2021automated, ruan2019deeplink, xie2019deeplink, 10.1016/j.infsof.2022.106961, 10546471}. Ruan et al.~\cite{ruan2019deeplink} introduced word embeddings and Recurrent Neural Networks (RNN) to capture semantic information from issues and commits. Similarly, Xie et al.~\cite{xie2019deeplink} employed neural networks for word embeddings and integrated knowledge graph mining to extract semantic context from code. However, these methods require large-scale data and do not apply to projects with poor quality or small scales.

\textcolor{black}{With advancements in Natural Language Processing (NLP), pre-trained language models (PLMs) like BERT and RoBERTa have been applied to software development tasks for artifact representation~\cite{batra2024esd}~\cite{nandwalkar2023descriptive}.  Existing fine-tuning-based ILR methods leverage the text understanding capabilities of PLMs by fine-tuning them to perform downstream ILR tasks. These pre-trained models are trained on large-scale, general corpora, giving them strong language understanding capabilities. For example, RoBERTa is trained using masked language model (MLM), while BERT incorporates both MLM and next sentence prediction during pre-training (as illustrated in Figure~\ref{figure: motivation}.a and Figure~\ref{figure: motivation}.b, respectively). Figure~\ref{figure: motivation}.c outlines the conventional fine-tuning approach for ILR, where PLMs serve as encoders for commits and issues. Additional neural network layers, such as fusion layers or fully connected layers, are often introduced, along with task-specific loss functions, to train the model effectively~\cite{lin2021traceability, lan2023btlink, Zhang2023EALinkAE}. However, a significant challenge with these methods arises from the mismatch between the pre-training tasks and the downstream ILR objectives. This misalignment can hinder the model’s ability to fully capture the semantic relationships between issues and commits, while also limiting the effective use of generalized knowledge embedded in PLMs~\cite{Zhang2023PromptLF,gu2021ppt}.}

Recently, a novel pre-training, prompting, and prediction paradigm, known as prompt-tuning, has achieved good performance in various tasks, including disease classification~\cite{yang2022knowledge}, relation extraction~\cite{chen2022knowprompt}, and news recommendation~\cite{Zhang2023PromptLF}. This new paradigm reorganizes downstream tasks into PLM tasks by prompting templates and label words~\cite{chen2022knowprompt, liu2023pre}. Inspired by this paradigm, we propose a \textbf{Multi-template Prompt-tuning} method with adversarial training for issue-commit \textbf{Link} recovery (MPLinker). \textcolor{black}{As shown in Figure~\ref{figure: motivation}.d, MPLinker integrates various templates to enhance the model's generalization and adopts adversarial training to mitigate the model overfitting.} The input to the model consists of commit, issue, templates (e.g., ``The link is $\left[MASK\right]$"), and label words (``correct", ``incorrect", etc.). The MLM in PLM then predicts ``$\left[MASK\right]$" to determine whether or not there are links. The following Research Questions (RQs) are proposed to investigate the practicality and effectiveness of the prompt-tuning paradigm in the ILR task:

\textbf{RQ1:} Which prompt-tuning architectures can perform better in MPLinker for the ILR task?

RQ1 compares the effectiveness of three prompt-tuning \textcolor{black}{architectures}, Single-template Prompt, Multi-template Prompt, and CLSPrompt in ILR. The experiment evaluated how well MPLinker performs ILR using MLM for cloze tasks. Additionally, the experiments confirmed that our proposed multi-template prompt can leverage the advantages of each template to improve overall ILR performance, reduce the occurrence of outliers, and achieve more stable and reliable results.

\textbf{RQ2:} Which PLMs are most effective for prompt-tuning in MPLinker for ILR task?

RQ2 compared RoBERTa, BERT, and GPT-2 to verify the importance of consistency between LLM training and downstream ILR tasks. The results indicate that RoBERTa, which utilizes MLM for cloze tasks during training, performs best among all LLMs. BERT  performs better than GPT-2, although it has less stability than GPT-2.

\textbf{RQ3:} How does adversarial training contribute to the MPLinker?

RQ3 evaluated the contribution of adversarial training to the performance of MPLinker in ILR through ablation experiments. The results demonstrate that incorporating adversarial training significantly improves the performance of ILR by enhancing \textcolor{black}{link} diversity and mitigating model overfitting.

\textbf{RQ4:} How effective is MPLinker in ILR compared to state-of-the-art methods?

RQ4 compared MPLinker with the state-of-the-art methods (FRLink~\cite{sun2017frlink}, DeepLink~\cite{ruan2019deeplink}, hybrid-linker~\cite{mazrae2021automated} and BTLink~\cite{lan2023btlink}) on the ILR task. The results show that MPLinker significantly enhances ILR performance, revealing prompt-tuning potential for ILR task.

In summary, the \textbf{main contributions} of this work are the following:
\begin{itemize}
\item  To the best of my knowledge, it is the first attempt to convert the ILR task into a cloze task through templates and propose a prompt-tuning-based ILR method. Moreover, this paper presents three prompt strategies for ILR and evaluates the impact of three PLMs on ILR performance.
\item MPLinker with multi-template is proposed to integrate predictions from various templates to determine whether issue and commit are linked. Since there is no prior knowledge about which template performs best, MPLinker with multi-template mitigates biases introduced by different templates, leading to fairer and more reliable results.
\item MPLinker incorporates adversarial training into the process by introducing noise into the text to perturb the issue and commit data, enhancing the model's robustness. Experimental results show that this strategy reduces model overfitting, thereby improving model performance in ILR task.
\end{itemize}

The rest of the paper is organized as follows:
Section \ref{Related work} introduces related work on fine-tuning and prompt-tuning PLMs for ILR. Section \ref{MPLinker} describes the MPLinker method with the three prompt-tuning \textcolor{black}{architectures} developed from various prompt strategies. Section \ref{Experiment} explains the experimental design, including datasets, evaluation measures, experiment setup, and baseline selections. The results of the experiment are presented and analyzed in Section \ref{Result and Analysis}. \textcolor{black}{Section \ref{disscussion} discusses the threats to the validity of the experiment and provide a usage guide for MPLinker. Section \ref{Conclusion} provides a summary of this work and outlines future works.}

\section{Related Work}
\label{Related work}
\textcolor{black}{This section discusses related work in the field of software traceability recovery. It also introduces the concept of prompt-tuning, along with relevant research in this area.}

\subsection{Software Traceability Recovery}
\textcolor{black}{
Early mainstream approaches to software traceability recovery were primarily based on information retrieval (IR) methods. These IR-based methods calculated text similarity by constructing retrieval models, such as the Vector Space Model \cite{hayes2006advancing}, Latent Semantic Indexing \cite{de2004enhancing}. Candidate traceability links were then identified by setting manual thresholds. To improve the performance of IR models, researchers introduced enhancements by incorporating various strategies, including text-based strategies \cite{gao2022using}, code structure-based strategies \cite{mahmoud2013supporting, gao2022propagating}, and model-based strategies \cite{rodriguez2020ir, rodriguez2020multi}.}

\textcolor{black}{
Machine learning-based methods build a large set of from traceability links, which serve as input data for training models. These models are then used to classify links as valid or invalid. Du et al. \cite{du2020automatic} trained a supervised model using features from IR models and query characteristics as inputs between different artifacts, then applied the trained model to  traceability recovery in the test datasets. SPLINT et al. \cite{dong2022semi} employed a semi-supervised learning approach to predict traceability links without labeled data. To reduce the amount of training data needed, ALCATRAL et al. \cite{mills2019tracing} integrated active learning with a supervised classifier. Comet et al. \cite{moran2020improving} used a probabilistic model to capture implicit relationships between developer feedback and groups of software artifacts.
}

\subsection{Fine-tuning PLMs for ILR}
Given a PLM for ILR, the previous fine-tuning-based approach is shown in Figure~\ref{figure: motivation}(a). First, the input commit and issue are converted into the input sequence of the PLM respectively (e.g. [CLS]... [SEP]). Then, the input sequences are encoded into corresponding output sequences through PLM, which is generally taken as the hidden layer of PLM. The spliced sequence is usually fed into a well-designed neural network to predict the probability value of label $y$ between issue and commit, resulting in a probability distribution of label $y$ on 0 and 1. If the probability value of label $y$ is higher for 0 than for 1, there is no link between the issue and the commit, and vice versa. Finally, the model parameters are usually fine-tuned using loss functions such as cross-entropy~\cite{lin2021traceability, lan2023btlink}.

In recent years, learning-based ILR methods have received attention from researchers. These methods perform similarity (e.g., Vector Space Model (VSM)~\cite{nguyen2012multi, sun2017frlink}) computation or classification (e.g., Random Forest~\cite{sun2017improving}, XGBoost~\cite{mazrae2021automated}) by constructing different feature inputs into the model. MLink~\cite{nguyen2012multi} used the issue description and the corresponding source code of the commit as its input. FRLink~\cite{sun2017frlink}  focuses on the code change documents and non-source documents in the commit and then calculates their similarity as features input to the VSM. PULink~\cite{sun2017improving} focused on unlabeled links as false samples. Then, the metadata and similarity features of the issue and commit are extracted as input to Random Forest for binary classification. The hybrid-linker~\cite{mazrae2021automated} employed text and non-text in artifacts, constructing an ensemble model that linearly combines classical classifiers. However, these approaches necessitate extensive feature engineering, and the feature selection needs researchers to have a substantial reserve of prior knowledge. This requirement not only increases the complexity and time required for model training but also introduces potential biases due to subjective feature selection.

There is a growing interest in developing approaches that minimize the need for manual feature engineering and leverage more automated techniques to enhance the performance of the ILR model. DeepLink~\cite{xie2019deeplink} used knowledge graphs for word embedding of code and employed the Word2Vec model for text embedding. The embedded words were then trained using a GRU to obtain feature representations, which were subsequently input into a Support Vector Machine (SVM) for binary classification of the links. Ruan et al.~\cite{ruan2019deeplink} proposed DeepLink, which employed \textcolor{black}{word2vec} embeddings~\cite{mikolov2013distributed} to represent features and subsequently fed them into an RNN~\cite{hochreiter1997RNN} for training. TraceNN ~\cite{guo2017semantically} is a tool that is similar to DeepLink, but its focus is on different artifacts. T-BERT~\cite{lin2021traceability} employed PLM for ILR. The model was initially pre-trained using CodeSearchNet and subsequently fine-tuned. \textcolor{black}{ Bai et al.~\cite{10.1016/j.infsof.2022.106961} proposed a time-based multi-dimensional recommendation approach that focuses on event data within GitHub, including artifacts such as commits, issues, pull requests, and pull request reviews. This method combines vector similarity, clustering techniques, and a deep learning model to improve the recommendation process. Additionally, Bai et al.~\cite{10546471} introduced a knowledge-aware heterogeneous graph learning method utilizing GloVe and random generation techniques for node embeddings. This approach defines eight distinct types of edge relationships, with the heterogeneous graph generating links through a metapath-based technique. However, this method primarily focuses on issues and pull requests within commit.} BTLink~\cite{lan2023btlink} employed PLMs (CodeBERT~\cite{feng2020codebert} and Roberta~\cite{liu2019roberta}) to represent programming language and natural language in artifacts separately. BTLink claimed they outperformed their compared methods (FRLink, DeepLink, and hybrid-linker) with deeper semantic understanding and better cross-project capabilities, representing it as the state-of-the-art method in ILR.

The current methods use PLMs to represent artifacts. However, the neural networks created for downstream ILR task do not align well with PLMs, leading to semantic information in PLMs not being fully utilised. To address this issue, the proposed approach uses prompt-tuning to transform the ILR task into a cloze task, which maintains consistency with PLMs through the ``$\left[MASK\right]$" prediction. 

\subsection{Prompt-tuning PLMs for ILR}
\begin{figure*}[h]

  \centering
  \includegraphics[width=1\linewidth]{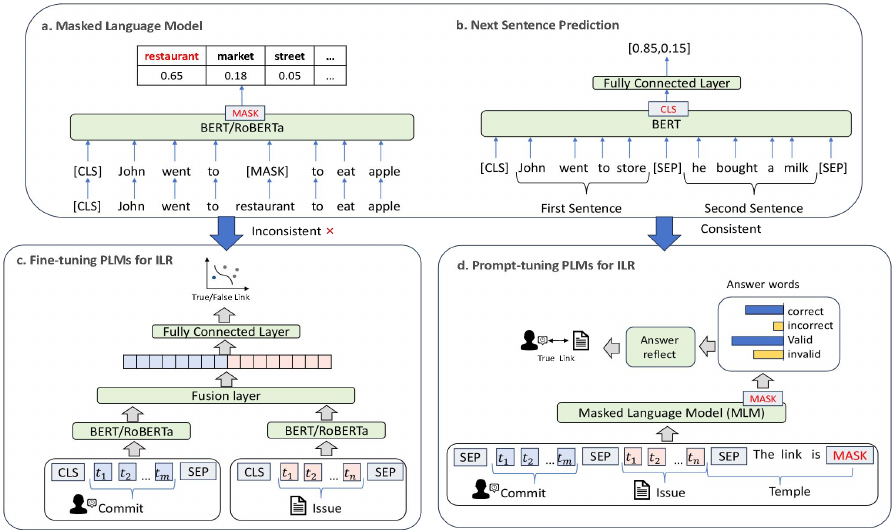}
  \caption{\textcolor{black}{\textcolor{black}{The motivation for prompt-tuning PLMs for ILR. Figure 1.a and Figure 1.b illustrate common pre-training tasks. Figure 1.a shows the masked language model (MLM), while Figure 1.b illustrates the next sentence prediction (NSP). Figures 1.c and 1.d depict the use of PLMs for downstream tasks. Figure 1.c shows the fine-tuning of the PLM for ILR, whereas Figure 1.d represents the prompt-tuning of the PLM for ILR.}}}
  \label{figure: motivation}
\end{figure*}

prompt-tuning is proposed mainly to bridge the gap between PLM and downstream tasks~\cite{chen2022knowprompt}. Unlike fine-tuning-based ILR methods, prompt-tuning requires the design of specific prompt templates that guide the PLM to generate the desired outputs (as shown in Figure~\ref{figure: motivation}(b))~\cite{li2021prefix}. This process involves creating a particular prompt template $x_{prompt}$ and a label set, then combining the issue, commit, and prompt template $x_{template}$ to form the input sequence for the PLM (e.g. [CLS] Issue [SEP] Commit [SEP] $x_{template}$). The number of ``[MASK]" tokens in the $x_{template}$ matches the number of predicted words, indicating the position of the label words (e.g. The link is [MASK]). The MLM layer in PLM predicts the right token for the ``[MASK]" positions, resulting in a probability distribution based on the PLM's vocabulary. This distribution is then mapped to the label set to obtain the probability values for each word. For instance, if the label set includes [``correct", ``incorrect"], the word with the highest probability value determines whether there is a link between the issue and commit.

Prompt-tuning has proven its methodological superiority in many domains. Chen et al.~\cite{chen2022knowprompt} were the first to apply prompt-tuning to relation extraction, integrating knowledge between relation labels into the prompt fine-tuning process to achieve knowledge-aware extraction. Yang et al.~\cite{yang2022knowledge} employed prompt-tuning for Automatic International Classification of Diseases (ICD) coding, achieving excellent results on few-shot medical datasets. Gu et al.~\cite{gu2021ppt} proposed a pre-training prompt-tuning method called PPT, exploring the effects of different prompt template combinations in few-shot scenarios. Their work demonstrated that combining soft and hard templates yielded impressive results in multiple general NLP tasks such as Sentence-Pair Classification, Multiple-Choice Classification, and Single-Sentence Classification. Some researchers have focused on prompt template design, with Gao et al.~\cite{gao2020making} being the first to propose automated methods for generating label sets and templates. Shin et al.~\cite{shin2020autoprompt} introduced a gradient-based approach to automatically generate words for label sets and templates. Furthermore, some studies have proposed continuous prompts~\cite{li2021prefix, he2021towards, hambardzumyan2021warp}, designing templates as learnable continuous embeddings instead of discrete words, and automating their training through models. These prompt-tuning methods align PLMs with downstream tasks through cloze tasks, effectively leveraging the textual information from the PLM pre-training process. Furthermore, by incorporating prompt templates to guide PLMs in completing various downstream tasks, these methods can be seamlessly adapted for ILR task. However, existing prompt-tuning methods do not apply to ILR.

\section{MPLinker}
\label{MPLinker}
\textcolor{black}{This section provides an overview of MPLinker, which consists of two main components: Prompt-tuning for ILR and Adversarial Training. The Prompt-tuning for ILR subsection includes three prompt-tuning architectures: Single-template Prompt, Multi-template Prompt, and CLSPrompt.}
 
\begin{figure*}[h]
  \centering
  \includegraphics[width=1\linewidth]{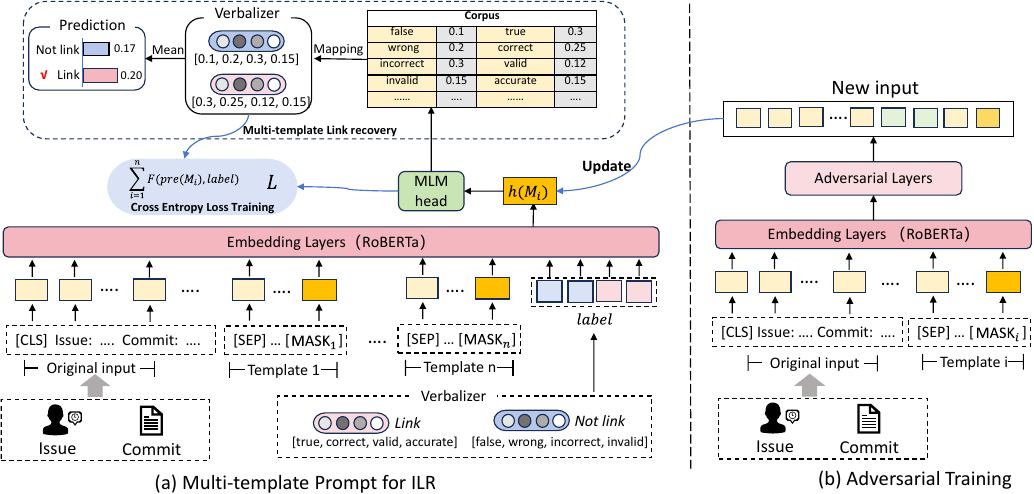}
  \caption{The overview of MPLinker. Multi-template Prompt on the left and adversarial training on the right.}
\label{overview}
\end{figure*}

 \textbf{Model input:} The model input data consists of $n$ commits $C=\{c_{1},c_{2}... ,c_{n}\}$ and $m$ issues $S=\{s_{1},s_{2}.... .s_{m}\}$ . For each $c_{i}$ in the commit, there is a corresponding $s_{j}$ to be linked, which is considered a true sample. The links other than $c_{j}$ are considered false samples. Figure~\ref{overview} shows a flowchart of the MPLinker process. The left side employs Multi-template Prompt (\textcolor{black}{see} Section~\ref{Multi-template Prompt}), while the right side introduces adversarial training to mitigate the model overfitting of the training process (\textcolor{black}{see} Section~\ref{Adversarial Training}). The details are explained as follows.

\subsection{Prompt-tuning for ILR}
\label{Prompt-tuning for ILR}
Different prompt strategies can lead to different  \textcolor{black}{prompt-tuning architectures}, making it essential to verify the effectiveness of the ``[MASK]" prediction and the proposed Multi-template Prompt method. As follows, we introduce three different Prompt-tuning \textcolor{black}{architectures} designed to enhance ILR: Single-template Prompt, Multi-template Prompt, and CLSPrompt. Instead of using the predicted probability of ``[MASK]" in MPLinker as a candidate answer, CLSPrompt obtains the predicted probability of ``[CLS]". The single-template Prompt uses one prompt word combined with MLM to guide the model in performing ILR, while the Multi-template Prompt integrates multiple single templates to obtain the final result.

\subsubsection{Single-template Prompt}
\label{Single-template Prompt}

\textbf{Input:} A prompt word consists of two components: a template and a set of label words. Consequently, we propose a prompt word construction method specifically tailored for the ILR task(as shown in Figure~\ref{overview2}(a)). Firstly, natural language and programming language, specifically the issue description, commit message, and commit code, are extracted from issues and commits to improving the model's comprehension. These extracted contents are referred to as $\left \langle ISSUE \right \rangle$ and  $\left \langle COMMIT \right \rangle$, respectively. The representation $\left \langle ISSUE \right \rangle$ and  $\left \langle COMMIT \right \rangle$ is defined as follows:
\begin{equation}
 <ISSUE> \leftarrow Issue: ID_{i}
\end{equation}

\begin{equation}
 <COMMIT> \leftarrow Commit: CM_{i} \left [SPE\right] CC_{i}
 \end{equation}

\textcolor{black}{Where $ID_{i}$ represents the issue description of $i$-th issue. $CM_{i}$ and $CC_{i}$ denote the commit message and commit code of $i$-th commit respectively. The prompt template $\left \langle TEMPLATE \right \rangle$ combines natural language components with a ``[MASK]” token, designed to guide the PLM model. This masked prediction facilitates completion of the downstream task of link recovery. To convert the ILR task into a cloze task, the input data is structured as follows:}

\begin{equation}
X_{prompt}=(<ISSUE>, <COMMIT>, <TEMPLATE>)
\end{equation}

Different templates can capture various features and patterns within the data, and each template is analyzed in Section~\ref{result 2}. \textcolor{black}{By framing the task as a cloze task, the model can predict the ``[MASK]” token]), thereby determining the link between the issue and commit.}

\textbf{Embedding Layers:} PLMs are trained on vast amounts of textual data, enabling them to capture rich semantic information. Applying these models for artifact embedding allows a deeper understanding of diverse text types and the generation of embeddings with profound semantic representation. Consequently, MPLinker utilizes pre-trained models from Hugging Face, which have been trained on extensive corpora, as its embedding layer. Through RQ2, the performance of three commonly used and representative \textcolor{black}{PLMs} in the field of ST was compared, leading to the selection of RoBERTa as the embedding layer for MPLinker.

The input sequence constructed from the issue and commit pairs is first passed through a PLM as embedding layer to generate contextual embeddings. Specifically, each token in 
 $\left \langle ISSUE \right \rangle$ and 
 $\left \langle COMMIT \right \rangle$ is mapped to its corresponding embedding vector using the PLM's embedding layer. \textcolor{black}{ Let $ \mathbf{E}_I$ denote the sequence of embeddings for the tokens in $\left \langle ISSUE \right \rangle$, and $\mathbf{E}_C$ represent the embeddings for  $\left \langle COMMIT \right \rangle$.   Additionally, let $\mathbf{E}_{T}$ denote the embedding of the prompt template. The overall embedding representation for the input sequence is created by concatenating the embeddings for the issue, commit, and template. The final embedding representation of the input sequence can be expressed as follows:}

\begin{equation}
\label{x_prompt_embedding}
\textcolor{black}{\mathbf{E}_{X_{prompt}}=\mathbf{E}_{I}\oplus\mathbf{E}_{C} \oplus \mathbf{E}_{T}}
\end{equation}

\begin{equation}
\textcolor{black}{\mathbf{E}_{T} = \mathbf{T}_{NL} \oplus \mathbf{h}_{M}}
\end{equation}
\textcolor{black}{in which $\oplus$ denotes the vector concatenation operation, $\mathbf{T}_{NL}$ represents  the embedding of natural language components and $\mathbf{h}_{M}$ is the embedding of the ``[MASK]" token.}

\textbf{Link Prediction:} 
\textcolor{black}{The word embedding $E_{X_{prompt}}$ is input into the MLM head, from which we extract the hidden layer $P_M$ corresponding to the $``[MASK]"$ token. It is important to note that the dimension of this hidden layer aligns with the model's vocabulary size, indicating that the output represents the probabilities of each word in the model’s vocab. Each pair of issue-commit ($\left \langle ISSUE \right \rangle$, $\left \langle COMMIT \right \rangle$) has a corresponding label $y$, categorized as linked (y=1) and not linked (y=0). Consequently, we define a verbalize $v(y)$ to map the probability space $P_M$ produced by MLM head to the ``[MASK]" token. }

\begin{equation}
\label{formula_v(y)}
v(y)=\left\{\begin{matrix}
  v_{pos}& y=1\\
  v_{neg}& y=0
\end{matrix}\right.
\end{equation}
\textcolor{black}{in which $v_{pos}$ indicates the answer space for artifacts with links (y=1), and $v_{neg}$ represents the answer space for artifacts without links (y=0). We then extract the probabilities corresponding to the words in $v_{pos}$ and $v_{neg}$  from the probability space, and then calculates their averages to obtain the label's probability distribution $pre(m)$. The formula is as follows:}


\begin{equation}
\label{formula_pre(m)}
\textcolor{black}{pre(m)=g(P_M([MASK]==v)|v \in v(y))}
\end{equation}
\textcolor{black}{in which $g(\cdot)$ denotes the function that converts the probability of model's vocabulary $P_M$ generated by the MLM head into the probability of label words $pre(m)$.}

\textbf{Training Model:} During training, the model updates its parameters to minimize the cross-entropy loss. This process enables the model to learn the mapping between the issue and commit sequences and the corresponding label, improving its ability to predict links in the ILR task accurately. This loss function is defined as follows:

\begin{equation}
L=-\frac{1}{N}\sum_{i=1}^{N}  [y_{i}\log_{}{pre_{i}(m)}  +(1-y_{i})log(1-pre_{i}(m))]
\end{equation}
\textcolor{black}{in which} $y_{i}$ and $pre_{i}(m)$ denote the gold label and model prediction probability of the input $i$-th issue and commit artifact.

\subsubsection{Multi-template Prompt}
\label{Multi-template Prompt}

\begin{figure*}[h]

  \centering
  \includegraphics[width=1\linewidth]{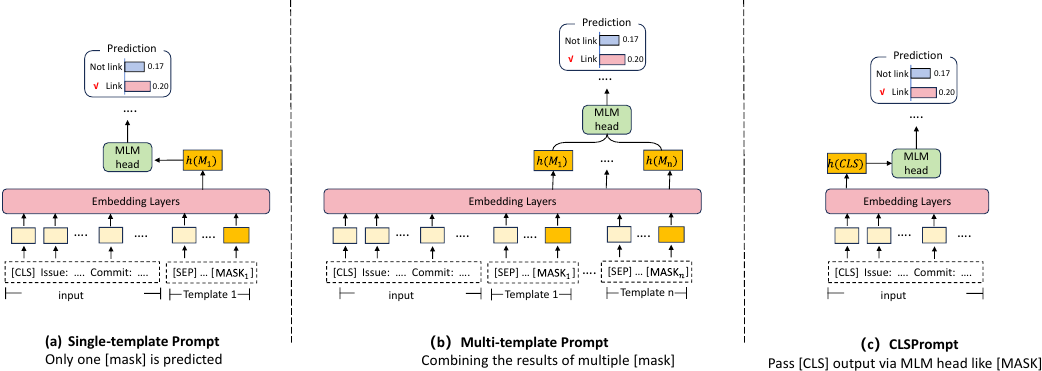}
  \caption{The comparison of Single-template Prompt, Multi-template Prompt, and CLSPrompt.}
\label{overview2}
\end{figure*}

Different templates may have distinct advantages and be suited to different dataset scenarios. Existing research lacks prior knowledge of which prompt template is most effective. Therefore, Multi-template Prompt calculates the final prediction by averaging the probability values of each single template result (as shown in Figure~\ref{overview2}(b)).

\textbf{Input:} In contrast to a single-template approach, the multi-template method involves constructing several prompt templates, each designed to capture different aspects of the data and enhance the model's ability to understand various patterns. For the ILR task, we construct multiple templates, each consisting of a combination of natural language components and a ``[MASK]" token, tailored to guide the PLM model in predicting the value of the ``[MASK]" token. These templates cover a broader range of potential relationships between issues and commits. \textcolor{black}{The set of prompt templates is denoted as $\left \langle T_1, T_2, \ldots, T_i, \dots  T_n \right \rangle$, $n$ represents the total number of templates and each template $T_i$ is designed to address different patterns and features.} The general structure of a multi-template input sequence can be expressed as:

\begin{equation}
X_{prompt}^{(i)} = (\langle ISSUE \rangle, \langle COMMIT \rangle, T_i)
\end{equation}


\textbf{Embedding Layers:} Similar to the single-template approach, the multi-template input sequences are passed through a PLM as an embedding layer to generate contextual embeddings. For each template $T_i$, the embedding representation is obtained as follows:
\begin{equation}
 \textcolor{black}{\mathbf{E}_{T_i} = \mathbf{T}_{NL_i}  \oplus \mathbf{h}_{M_i}}
\end{equation}
\textcolor{black}{in which $\mathbf{T}_{NL_i}$ is the embedding of the natural language components of the $i$-th template and $\mathbf{h}_{m_i}$ is the embedding of the $i$-th ``[MASK]" token.} The combined embedding for the input sequence using the $i$-th template is:

\begin{equation}
\label{equation_11}
\mathbf{E}_{X_{prompt}}^{(i)} = \mathbf{E}_{I} \oplus \mathbf{E}_{C} \oplus \mathbf{E}_{T_i}
\end{equation}

\textbf{Link Predictions:} 
\textcolor{black}{Similar to the single-template prompt, the word embedding $\mathbf{E}_{X_{prompt}}^{(i)}$ corresponding to the $i$-th template is input into the MLM head, and the hidden layer corresponding to the ``[MASK]" token is extracted to obtain the probability distribution over the model's vocabulary, denoted as $P_{M_i}$. Reference to Equations \ref{formula_v(y)} and \ref{formula_pre(m)}, the vocabulary probabilities $P_{M_i}$ are converted into the probability for the label words $pre(m_i)$. Subsequently, we aggregate these predictions to determine the final output. The aggregated probability can be calculated as follows:}

\begin{equation}
\label{pre_equation}
pre(m)=\frac{1}{n}  {\textstyle \sum_{i=1}^{n}} pre(m_{i})
\end{equation}
By averaging the predictions from multiple templates, the model can leverage the strengths of each template, capturing different aspects of the input data and providing a more robust and comprehensive prediction.

\textbf{Training Model:} Similarly,
the model fine-tunes its parameters using the cross-entropy loss function. This allows the model to better learn the mappings between issue and commit sequences and their corresponding labels.

\subsubsection{CLSPrompt}
As shown in Figure~\ref{overview2}(c), CLSPrompt enhances the previous \textcolor{black}{architectures} by eliminating the use of the \textcolor{black}{MLM} for predicting the ``[MASK]" token. Instead, it leverages the probabilities associated with the ``[CLS]" token, which is traditionally used in classification tasks within pre-trained language models (PLMs).

\textbf{Input and Embedding Layers:} Similar to the single-template and multi-template approaches, the input sequence in CLSPrompt is constructed from the $\langle ISSUE \rangle$ and $\langle COMMIT \rangle$ components, along with a template. However, instead of focusing on predicting a masked token, the entire sequence is processed through the PLM. The embedding corresponding to the ``[CLS]" token is used for final prediction. Let $\mathbf{E}_{CLS}$ denote the embedding of the ``[CLS]" token, which is the first token of the input sequence and typically represents the overall summary of the sequence. \textcolor{black}{The word embedding for the input sequence is obtained as:}

\begin{equation}
\mathbf{E}_{X_{prompt}} = \mathbf{E}_{I} \oplus \mathbf{E}_{C} \oplus  \mathbf{E}_{CLS}
\end{equation}
\textcolor{black}{in which $\mathbf{E}_{I}$ and $\mathbf{E}_{C}$ are the embedding for the issue and commit respectively. The $\oplus$ denotes vector concatenation.}

\textbf{Link Prediction:} Unlike the previous methods, CLSPrompt does not rely on the MLM head to predict the ``[MASK]" token. Instead, the probability distribution over the potential candidate labels (linked or not linked) is directly derived from the output corresponding to the ``[CLS]" token. The predicted probability $pre_{CLS}(c)$ for a candidate label $c$ is given by:

\begin{equation}
pre_{CLS}(c) = p(h_{CLS} = v(c) | \mathbf{E}_{X_{prompt}})
\end{equation}
\textcolor{black}{in which} $h_{CLS}$ represents the output embedding of the ``[CLS]" token, and $v(c)$ is the verbalize mapping the model's output to the label space.

\textbf{Training Model:} CLSPrompt fine-tunes the model by optimizing the cross-entropy loss function. Instead of predicting masked tokens, the model learns to classify sequences directly based on the ``[CLS]" token's output. The loss function is defined as:

\begin{equation}
L=-\frac{1}{N}\sum_{i=1}^{N}  [y_{i}\log_{}{pre_{CLS_{i}}(c)}  +(1-y_{i})\log(1-pre_{CLS_{i}}(c))]
\end{equation}
\textcolor{black}{in which $y_{i}$ represents the true label, and $pre_{CLS_{i}}(c)$ indicates the predicted probability for the $i$-th input sequence. In summary, CLSPrompt offer a streamlined approach that directly utilizes the output of the ``[CLS]" token.}

\textcolor{black}{The effectiveness of three different prompt architectures for ILR were evaluated (see ~\ref{Result 1}). The experiment shows that Multi-template prompt is the most effective, which is used as the prompt-tuning method for MPLinker.}

\subsection{Adversarial Training}
\label{Adversarial Training}

Adversarial training involves introducing intentionally designed disturbances during the training phase, allowing the model to better cope with noise and anomalies in the data. It enhances the robustness of the model in ILR, enabling it to accurately ILR amidst data uncertainties. As shown in Figure~\ref{overview}(b), MPLinker incorporates adversarial training to diversify the training data, thereby reducing the model overfitting.

Adversarial training in MPLinker consists of generating text perturbations and calculating cross-entropy loss. We employ the Projected Gradient Descent (PGD)~\cite{zhu2019freelb} to create these perturbations and use the \textcolor{black}{cross-entropy} loss function to evaluate their impact. PGD is commonly used in text perturbations~\cite{bayer2022survey, geisler2024attacking, moothedath2023comparing}. \textcolor{black}{
In this section, we use the single-template prompt as an example. As illustrated in Equation \ref{x_prompt_embedding}, the  PLM serves as the embedding layer to generate the word embeddings $\mathbf{E}_{X{prompt}}$ and the corresponding label $y$. An initial perturbation is then applied to the embedding vector $\mathbf{E}_{X{prompt}}$, defined as follows:}

\begin{equation}
\delta^{(0)} \sim \mathcal{U}(-\epsilon, \epsilon)
\end{equation}
\textcolor{black}{in which $\delta^{(0)}$ represents the perturbation matrix generated in the 0-th iteration, while $\mathcal{U}(\cdot)$ denotes a uniform distribution function that generates the initial perturbation. Specifically, the function generates a matrix of the same dimensions as the input embedding vector $\mathbf{E}_{X{prompt}}$, with each element randomly sampled from the range $[-\epsilon, \epsilon]$. For this study, $-\epsilon$ is set to 1 by default. The perturbation matrix is then added to the input embedding vector $\mathbf{E}_{X{prompt}}$, and the initial adversarial vector ${E}_{X_{adv}} ^{0}$, as shown in the following equation:}

\begin{equation}
\label{equation_17}
\textcolor{black}{{\mathbf{E}}_{X_{adv}} ^{0} = {\mathbf{E}}_{X_{prompt}}^{(i)}  + \delta^{(0)}}
\end{equation}

\textcolor{black}{We incorporate adversarial learning to jointly fine-tune the PLM. Using the single-template prompt for training as an example, we replace the word embedding $\mathbf{E}_{X{prompt}}$ in Section \ref{Prompt-tuning for ILR} with the adversarial embedding $\mathbf{E}_{X_{adv}} ^{0}$. Similar to Equation \ref{pre_equation}, the MLM predict the ``[MASK]" token to obtain $pre(m)$, followed by calculating the cross-entropy loss between this prediction and the true label $y$. This loss is then backpropagated to fine-tune the PLM. The formulation is as follows:}

\begin{equation}
\label{equation_18}
g^{(t)} = \nabla_\delta \mathcal{L}(X + \delta^{(t)}, y)
\end{equation}
\textcolor{black}{in which $y$ represents the label associated with the perturbed vector $\mathbf{E}_{X_{adv}} ^{0}$.} The perturbation $ \delta$ is then updated using $ L_\infty$ norm constraint gradient:

\begin{equation}
\label{equation_19}
\delta^{(t+1)} = Clip_{[-\epsilon, \epsilon]} \left( \delta^{(t)} + \alpha \cdot \text{sign}(g^{(t)}) \right)
\end{equation}
\textcolor{black}{ in which $\delta^{(t+1)}$ denotes the perturbation matrix at the $(t+1)$-th iteration, and $\delta^{(t)}$ indicates  the perturbation matrix at the $t$-th iteration. The $sign(\cdot)$ function produces a matrix with values of either 1 or -1, matching  in dimension and sign of the perturbation matrix $\mathbf{E}_{X_{adv}}^{0}$. Here, $\alpha$ represents the step size of the perturbation, with larger values generating stronger perturbations. In our experiment, $\alpha$ is set to 1 by default.  $Clip_{[-\epsilon, \epsilon]}(\cdot)$ ensures that each element of the perturbation remains within the range $([- \epsilon, \epsilon])$. Specifically, if any element of the perturbation matrix is less than $- \epsilon$, it is set to  $- \epsilon$, and if any element exceeds $\epsilon$,  it is set to $\epsilon$.}

\textcolor{black}{This process is repeated from Equations \ref{equation_17} to \ref{equation_19} over $T$ iterations to obtain the final adversarial vector. In this study, we generate the perturbation for each input only once, setting $T$ to 1. After $T$ iterations, the final adversarial example is expressed as follows:}

\begin{equation}
{E}_{adv} ^{T} = {E}_{X_{prompt}}  + \delta^{(T)}
\end{equation}

\section{Experiment}
\label{Experiment}
\textcolor{black}{This section offers an overview of the datasets used in the experiments and the evaluation metrics employed to assess the model's performance. Additionally, it also outlines the experimental setup and the selection of baselines.}

\subsection{Datasets}
\textcolor{black}{To comprehensively evaluate MPLinker, the experiments are conducted on six open-source projects sourced from GitHub~\cite{ZHANG2022106797, Bai2023AutomatingDS}, as used in previous works~\cite{lin2021traceability, lan2023btlink}: log4net, OODT, Giraph, Keras, Nutch and Isis.} These projects differ in size, language, and source. The experiment divided each dataset into train, valid, and test sets in an 8:1:1 ratio. The dataset is available at \url{https://figshare.com/s/f5d4a244388581a14e5c}.

\begin{table}[htbp]
\centering
\caption{The details of experimental datasets.}
\label{The details of experimental datasets.}
\footnotesize
\begin{tabular}{cccccc}
\hline
Project & Issues & Commits   & True links  & Flase links & Language   \\
 \hline
log4net  & 239&	115 &	266&	866&	C\#  \\
OODT & 727 &2251	 &	1329&9072	&	Java, Python \\
Giraph  &	915 &1138	&878	&8298&Java	 \\
keras & 717	&719&		725& 13999&		 Python \\
Nutch &1820 &3280	 &1821	&11210	&Java	 \\
Isis &2159 &18864	 &12700	&129435	&	Java \\
\hline
\end{tabular}
\end{table}

\subsection{Evaluation Measures}
To comprehensively evaluate the model, the experiments compared the performance of the methods across six metrics. In addition to the commonly used measures in the ST, such as Precision, Recall, and F1-score. \textbf{Precision} measures the accuracy of the identified links between issues and commits. \textbf{Recall} measures the ability of the model to identify all relevant issue-commit links correctly. Their formulas are as follows:

\begin{eqnarray}
Precision = \frac{TP}{TP + FP}
\end{eqnarray}

$TP$ is the number of correctly identified issue-commit links. $FP$ represents the number of incorrectly identified issue-commit links.

\begin{eqnarray}
Recall = \frac{TP}{TP + FN} 
\end{eqnarray}

$FN$ is the number of relevant issue-commit links that were not identified.

Recall and Precision have an inverse relationship; an increase in recall may lead to decreased precision. To balance the trade-off between the two, the \textbf{F1-score} is also used for evaluation in the experiment. 

\begin{align}
 F1-score = 2 \times \frac{Precision \times Recall}{Precision + Recall} 
\end{align}

The experiments also utilized MCC (Matthews Correlation Coefficient), ACC (Accuracy), and AUC (Area Under the ROC Curve), which are standard measures for assessing classification models. \textbf{AUC} represents the area under the Receiver Operating Characteristic (ROC) curve, which plots the true positive rate against the false positive rate at various threshold settings. The formulas for \textbf{MCC} and \textbf{ACC} are as follows.

 
\begin{footnotesize} 
\begin{align}
MCC = \frac{TP \times TN - FP \times FN}{\sqrt{(TP + FP)(TP + FN)(TN + FP)(TN + FN)}} 
\end{align}
\end{footnotesize} 

Where $TN$ represents the number of false links correctly predicted as negative.

\begin{align}
ACC = \frac{TP+FN}{TP+FN+FP+TN} 
\end{align}

In analyzing the experimental results, two non-parametric tests are used: the Wilcoxon signed-rank test ~\cite{woolson2005wilcoxon} and Cliff’s Delta effect size test ~\cite{macbeth2011cliff} to compare MPLinker with other methods. The Wilcoxon signed-rank test is a non-parametric statistical method for small sample sizes. It is employed to compare whether there is a significant difference between two strategies or methods across all datasets. If the calculated p-value is less than 0.05, it indicates a significant difference between the two samples, meaning that one method significantly outperforms the other ~\cite{kitani2022one}. Cliff's Delta is a measure of effect size used to assess the difference between two independent samples. If Cliff's Delta value $\ge $ 0.33, indicating a significant difference between the samples ~\cite{meissel2024using}.

\subsection{Experiment Setup}
 The training and validation datasets are fed into the model in the prompt-tuning stage. The training set is used to adjust the model parameters, while the validation set assesses the performance at the end of each epoch. The initial learning rate and the weight decay are 0.01,  and the batch size is 8. The maximum input sequence length for the PLM is set to 512. The model is trained for 20 epochs, and the one that achieves the highest accuracy on the validation set is chosen as the final model. This model is then tested using the test set.

\subsection{Baselines}
This paper selects four representative ILR methods as baselines for comparison with MPLinker's performance. \textcolor{black}{These baselines include the latest ILR methods based on machine learning and deep learning. The comparison effectively highlights MPLinker's performance in ILR.}

\textbf{FRLink}~\cite{sun2017frlink} focuses on non-source documents, uses TF-IDF for artifact representations, and then calculates code similarity and text similarity. The final threshold is determined by the threshold learning algorithm, and a similarity greater than the threshold is selected as the link prediction result.

\textbf{DeepLink\footnote{https://github.com/ruanhang1993/Deeplink}} ~\cite{ruan2019deeplink} DeepLink was the first to use deep learning as a representation method for artifacts. In this approach, textual artifacts and code artifacts are separately embedded using word2vec and then input into an LSTM for representation. Finally, link recovery is performed using cosine similarity.

\textbf{hybrid-linker\footnote{https://github.com/MalihehIzadi/hybrid-linker}} ~\cite{mazrae2021automated}  is an ensemble approach based on machine learning. The link recovery results are obtained by linearly combining different classifiers for textual and non-textual prediction.

\textbf{BTLink\footnote{https://github.com/OpenSELab/BTLink/} (2023)} ~\cite{lan2023btlink} is a state-of-the-art ILR method based on the Bert framework. This method uses the CodeBERT and Roberta models as the word embedding layer to obtain feature representations of issue text, commit text, and commit code. Then, it constructs a fusion layer for the issue text in combination with the commit code and commit text.

\section{Result and Analysis}
\label{Result and Analysis}
This section evaluates MPLinker and presents experimental results by answering the following four research questions. RQ1 explores different prompt strategies to determine the optimal prompt-tuning framework. RQ2 investigates the three most common PLMs to select the best one for artifact representation and mask prediction. RQ3 assesses the contribution of adversarial training to MPLinker, and RQ4 compares the MPLinker with the state-of-the-art ILR methods.

\subsection{\textcolor{black}{Effectiveness of Prompt-tuning Architectures (The result for RQ1)}}
\label{Result 1}

\begin{table}
\scriptsize
\centering
\caption{The content of Single-template Prompt. }
\label{result RQ1_1}
\begin{tabular}{cp{7cm}}
\hline
\textbf{Template} &\textbf{Content}\\
\hline
 \textbf{Single-template Prompt 1}& [SEP] The link is [MASK]
 \\
\textbf{Single-template Prompt 2}&
[SEP] The answer between issue and commit is [MASK]
\\ \textbf{Single-template Prompt 3}&
[SEP] Does the commit answer the issue? [MASK]
\\
 \hline
\end{tabular}
\end{table}

\begin{table}
\scriptsize
\centering
\setlength{\tabcolsep}{2pt}
\caption{\textcolor{black}{Performance Comparison of Single-template Prompt (ST Prompt), Multi-template Prompt (MT Prompt), and CLSPrompt for ILR.}}
\label{result RQ1_2}
\begin{tabular}{cccccccc}
\hline
\multirow{2}{*}{\textbf{Project}} &\multirow{2}{*}{\textbf{Template}}  && &\textbf{Measures}&& \\

& & \textbf{F1} & \textbf{Precision} & \textbf{Recall} & \textbf{MCC} & \textbf{AUC} & \textbf{ACC} \\
 \hline
\multirow{5}{*}{\textbf{log4net}} & \textbf{ST Prompt 1}
& 84.62 &	95.65 &	75.86 &	\textcolor{red}{89.64} &	87.27& 	92.45 
 \\
 &\textbf{ST Prompt 2}&85.19 &	92.00& 	\textcolor{red}{79.31} 	&80.56 &	88.36& 	92.45 
\\
& \textbf{ST Prompt 3}&81.48 &	88.00& 	75.86 &	75.57 &	85.98 &	90.57
 \\
 & \textbf{MT Prompt}&
 
 \textcolor{red}{86.79} &	\textcolor{red}{96.83} &	\textcolor{red}{79.31} &	86.79& 	86.79 &	\textcolor{red}{93.40}  
 \\

 &\textbf{CLSPrompt} &86.45& 	95.00& 	\textcolor{red}{79.31} 	&85.78 	&\textcolor{red}{89.66}& 	88.46 \\	

\hline
\multirow{5}{*}{\textbf{OODT}} & \textbf{ST Prompt 1}
& 82.70 &	85.96 	&79.67 	&80.43& 	88.92 	&95.86 
 \\
 &\textbf{ST Prompt 2}&83.56 &	\textcolor{red}{92.16} &	76.42 &	81.92 	&87.75 	&96.27 
\\
& \textbf{ST Prompt 3}&84.26 &	88.39 &	80.49 &	82.26 	&89.50& 	96.27 
 \\
 & \textbf{MT Prompt}&\textcolor{red}{86.32} &	90.99 &	\textcolor{red}{82.11} &	\textcolor{red}{84.64}& 	\textcolor{red}{90.48} &	\textcolor{red}{96.77} 
 \\
  &\textbf{CLSPrompt} &81.82 &	83.19& 	80.49 &	79.30 &	89.09& 	95.56 
 \\	
 \hline
\multirow{5}{*}{\textbf{Giraph}} & \textbf{ST Prompt 1}
& 87.74 &	95.77 &	80.95 &	86.70 &	90.25 &	86.74 
 \\
 &\textbf{ST Prompt 2}&82.08& 	91.30 	&74.56& 	80.21 &	86.07 &	95.73 
\\
& \textbf{ST Prompt 3}&88.31 &97.14 &	\textcolor{red}{80.95} 	&87.42 &	90.32 &	97.57 
 \\
 & \textbf{MT Prompt}&\textcolor{red}{88.89} &	\textcolor{red}{98.55} &	\textcolor{red}{80.95} &	\textcolor{red}{88.15} &	\textcolor{red}{90.40} &	\textcolor{red}{97.71} 
\\
   &\textbf{CLSPrompt} &86.62 	&93.15& 	\textcolor{red}{80.95}& 	85.31 &	90.10 &	97.17

 \\
 \hline
 \multirow{5}{*}{\textbf{Keras}} & \textbf{ST Prompt 1}
& 88.24 &	88.24 &	88.24& 	72.33 &	86.11 &	86.11 
 \\
 &\textbf{ST Prompt 2}&91.03& 	90.97 &	\textcolor{red}{91.67}& 	90.97& 	90.97 &	90.97 
\\
& \textbf{ST Prompt 3}&90.65 &	94.03 &	87.50 &	82.14 	&90.97& 	90.97 
 \\

 & \textbf{MT Prompt}&\textcolor{red}{91.55} &	\textcolor{red}{92.86} &	90.28 &	\textcolor{red}{91.67} &	\textcolor{red}{91.67} 	&\textcolor{red}{91.67} 
 \\

   &\textbf{CLSPrompt} &81.82 &	83.19& 	80.49 &	79.30& 	89.09& 	95.56 

 \\
 \hline

   \multirow{5}{*}{\textbf{Nutch}} & \textbf{ST Prompt 1}
& 80.00 	&89.71 	&72.19 &	77.93& 	85.47 	&95.26 

 \\
 &\textbf{ST Prompt 2}&81.70 &	91.24 &	73.96 &	79.82 &	86.45 	&95.65

\\
& \textbf{ST Prompt 3}&81.79 &	88.89 &	\textbf{75.74} 	&79.62 &	87.15 &	95.57 

 \\
 & \textbf{MT Prompt}&\textcolor{red}{82.08} &	\textcolor{red}{91.30} 	&74.56&	80.21 &	\textcolor{red}{86.74} 	&\textcolor{red}{95.73} 

 \\
    &\textbf{CLSPrompt} &73.37 &	91.85& 	73.37& 	\textcolor{red}{86.19} &	79.79 &	95.65 

    \\
 \hline

  \multirow{5}{*}{\textbf{Isis}} & \textbf{ST Prompt 1}
& 86.67 &	92.51 	&81.53 &	85.57 &	90.42 &	97.61 
 \\
 &\textbf{ST Prompt 2}&86.92 &	\textcolor{red}{95.42} &	79.81 	&86.09 &	89.70 &	97.71 
\\
& \textbf{ST Prompt 3}&88.58 	&94.28 &	83.54 &	87.65& 	\textcolor{red}{91.50} &	97.94 
 \\
 & \textbf{MT Prompt}&\textcolor{red}{88.90 }&	93.09 &	\textcolor{red}{85.07} 	&\textcolor{red}{92.20}& 	87.89& 	\textcolor{red}{97.97} 
 \\
     &\textbf{CLSPrompt} &88.55& 	94.08 &	83.64 &	87.60 &	91.54 &	97.94 
    \\
 \hline
\end{tabular}
\end{table}

\begin{table}
\centering
\scriptsize
\setlength{\tabcolsep}{3pt}
\caption{\textcolor{black}{\textcolor{black}{Statistical analysis of Multi-template Prompt (MT Prompt) with Single-template  Prompt (ST Prompt) and CLSPrompt using Wilcoxon Signed-Rank Tests. The * indicate significance levels of 0.05. The p-value $<$ 0.05 indicates a statistically significant difference between the two group.}}}
\label{result RQ1_3}
\begin{tabular}{lcccccc}
\hline
\multirow{2}{*}{\textbf{Approaches}}  && &\textbf{p-value}&& \\
& \textbf{F1} & \textbf{Precision} & \textbf{Recall} & \textbf{MCC} & \textbf{AUC} & \textbf{ACC} \\
\hline
\multirow{1}{*}{\textbf{MT Prompt vs ST Prompt 1} } 
&0.0313*& 	0.0313*& 	0.0431* &	0.1563 &	0.5625& 	0.0313* 
\\
\multirow{1}{*}{\textbf{MT Prompt vs ST Prompt 2}} &0.0313* &	0.4375& 	0.1380& 	0.0313* &	0.5625&	0.0313*
\\
\multirow{1}{*}{\textbf{MT Prompt vs ST Prompt 3}}&		0.0313* &	0.1563 &	0.0796& 	0.0313* 	&0.6875 	&0.0313* 
\\
\multirow{1}{*}{\textbf{MT Prompt vs  CLSPrompt}}  &	0.0313* &	0.1563 	&0.1088& 	0.3125 &	0.8438 &	0.0313* 
\\
 \hline
\end{tabular}
\end{table}

\textbf{Approach.}
Experiments evaluated the performance of different prompt-tuning \textcolor{black}{architectures} on the same dataset to ensure differences solely in model architecture. For a single prompt template, three types of prompts were designed(as shown in Table~\ref{result RQ1_1}) to explore the impact of templates on ILR. For a comprehensive evaluation, six performance metrics (including \textcolor{black}{F1-score}, Precision, Recall, MCC, AUC, and ACC) were used to validate the effectiveness across six datasets. The Wilcoxon signed-rank test was also employed to assess the significance of differences among the three \textcolor{black}{architectures} statistically.

\textbf{Result.}
Table~\ref{result RQ1_2} shows the performance of the three \textcolor{black}{prompt-tuning architectures}, with the highest values for each metric highlighted in red. Table~\ref{result RQ1_3} presents Wilcoxon signed-rank test results comparing multi-template prompts with other \textcolor{black}{architectures}. \textbf{While differences among the three models are minor, the Multi-template Prompt showed the best overall performance.} 
Specifically, it outperformed the average of the three single-template prompts by 2.12\% in \textcolor{black}{F1-score}, 2.18\% in Precision, 2.14\% in Recall, 4.67\% in MCC, 0.49\% in AUC, and 1.56\% in ACC. Compared to the CLSPrompt, it improved by 2.8\% in F1, 2.47\% in Precision, 0.71\% in Recall, 2.92\% in MCC, 0.47\% in AUC, and 1.25\% in ACC. Additionally, there are no consistent results across six projects indicating which single-template content is best, with some showing negative effects, such as single-template prompt 2 performing worse in Isis with an F1-score lower than CLSPrompt (86.92\%/88.58\%).  These findings highlights the importance of prompt design for ILR task. Multi-template prompts average the results of multiple single-template results, smoothing the predictions and reducing the occurrence of extreme predictions for a single template, making the final results more stable and credible. Fortunately, the proposed Multi-template prompt method balances the strengths and weaknesses of single templates, achieving complementary effects. Although improvements are modest, significantly surpassing other \textcolor{black}{architectures} only in F1-score and ACC, its comprehensive metrics across all projects demonstrate its strong generalizability. \textbf{ In the subsequent experiments, MPLinker will default to Multi-template Prompt for the final decision.}

\subsection{\textcolor{black}{Comparison of PLMs for Prompt-tuning (The result for RQ2)}}
\label{result 2}

\begin{figure*}[h]
  \centering
  \includegraphics[width=1\linewidth]{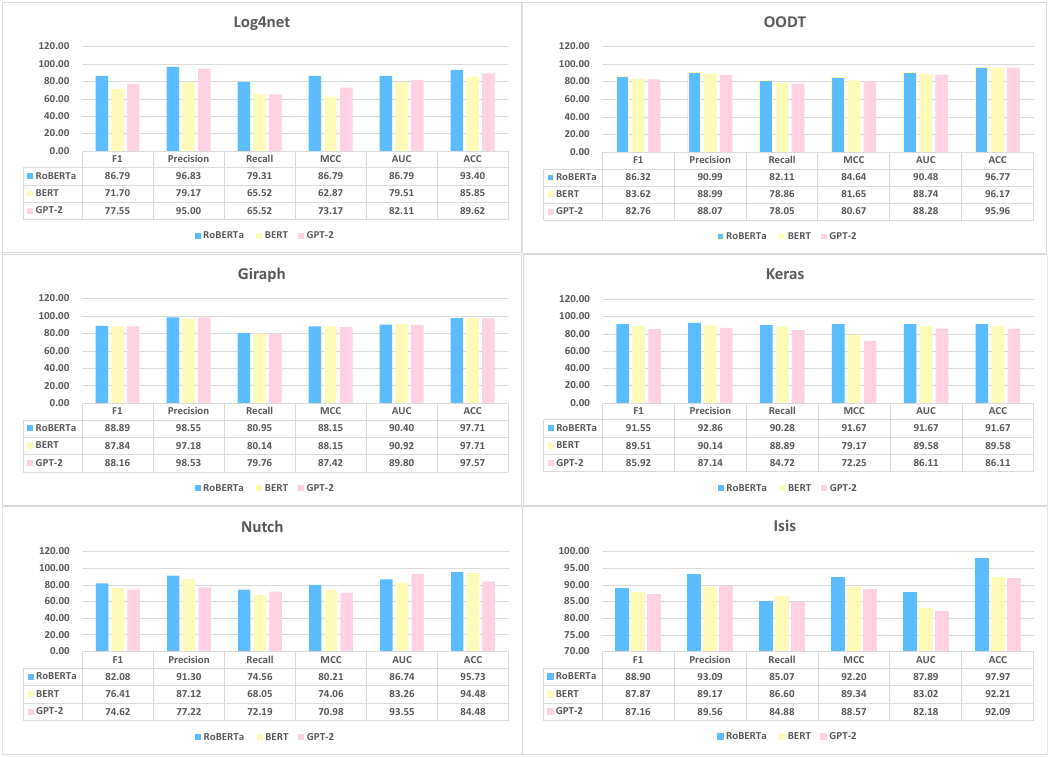}
  \caption{Performance comparison of three PLMs across six projects.}
\label{result RQ2_pci_1}
\end{figure*}

\begin{figure*}[h]
  \centering
  \includegraphics[width=1\linewidth]{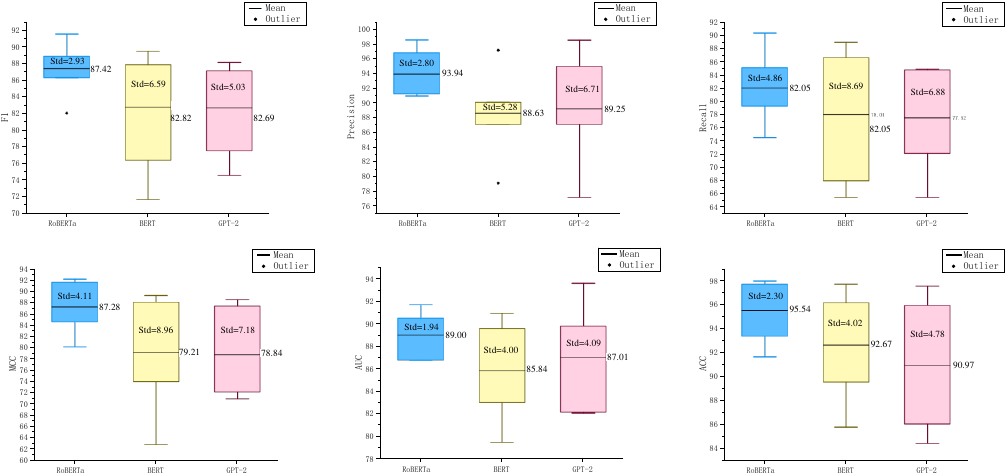}
  \caption{The mean, standard deviation for the ILR task across three PLMs: RoBERTa, BERT, and GPT-2.}
  \label{result RQ2_pci_2}
\end{figure*}

\begin{table}[htbp]
\centering
\scriptsize
\setlength{\tabcolsep}{2pt}
\caption{\textcolor{black}{The results of Wilcoxon signed-rank tests and Cliff’s Delta hypothesis tests for the ILR task across three PLMs: RoBERTa, BERT, and GPT-2. The $\ast$ indicate significance levels of 0.05.}}
\label{result RQ2_2}
\begin{tabular}{cccccccc}
\hline
\multirow{2}{*}{\textbf{Project}} &\multirow{2}{*}{\textbf{PLMs}}  && &\textbf{Measures}&& \\
& & \textbf{F1} & \textbf{Precision} & \textbf{Recall} & \textbf{MCC} & \textbf{AUC} & \textbf{ACC} \\
 \hline
\multirow{2}{*}{\textbf{p-value}} &\textbf{RoBERTa vs BERT}& 0.0625& 	0.0313 *&	0.0625& 	0.0431*	&0.0625 &	0.0431*
\\
&\textbf{RoBERTa vs GPT-2 }& 0.0313* &	0.0313* &	0.0313* &	0.0313*& 	0.4375 &	0.0313* 
\\
\hline
\multirow{2}{*}{\textbf{Cliff's Delta}} &\textbf{RoBERTa vs BERT}&0.2222 &	0.7222 &	0.3333 & 	0.5278 &	0.3889 &	0.4167 
\\
&\textbf{RoBERTa vs GPT-2}&0.5556 &	1.0000& 	0.3889 &	0.5556 &	0.3333 &	0.5556
\\
 \hline
 
\end{tabular}
\end{table}

\textbf{Motivation.} 
MPLinker encodes issues and commits using PLMs, predicting label probabilities through the MLM layer. Thus, exploring the effects of different PLMs on MPLinker for ILR is essential. Additionally, RQ2 experimentally verifies the impact of inconsistencies between PLMs and downstream ILR task on model performance.

\textbf{Approach.} Experiments were conducted on six projects, using six metrics for comprehensive evaluation. Besides analysing commonly used MLM-based PLMs like RoBERTa\footnote{https://huggingface.co/FacebookAI/roberta-base}~\cite{roberta} and BERT\footnote{https://huggingface.co/google-bert/bert-base-uncased}~\cite{bert}, we also assess the performance of the autoregressive model called GPT-2\footnote{https://huggingface.co/openai-community/gpt2}~\cite{gpt2}. Specifically, the experiment involved appending prompt templates to the input sequence of GPT-2, which expands the word embedding layer in the PLM. These PLMs generate rich contextual embeddings that capture deep semantic and syntactic information for software artifacts and have demonstrated strong performance across a wide range of natural language processing tasks. To assess the generalization and overall performance of the three PLMs, mean scores across all metrics were calculated. The Wilcoxon signed-rank tests and Cliff’s Delta were employed to compare the significant differences among the PLMs. Finally, the standard deviation was used to assess the stability of the PLMs.

\textbf{Result.} Figure~\ref{result RQ2_pci_1} presents the performance of each PLM on six projects. Figure~\ref{result RQ2_pci_2} shows the stability of the three PLMs in MPLinker through box plots. Table~\ref{result RQ2_2} shows the statistical analysis results of the PLMs across six metrics. \textcolor{black}{In comparisons of individual datasets  (as shown in Figure~\ref{result RQ2_pci_1}), RoBERTa performed best in four projects: log4net, OODT, Giraph and Keras.} In Isis, RoBERTa's Recall decreased by 1.53\% compared to BERT, but its \textcolor{black}{F1-score} and Precision increased by 1.03\% and 3.92\%, respectively. Therefore, \textcolor{black}{a slight sacrifice in} Recall to improve \textcolor{black}{F1-score} and Precision is acceptable. In Nutch, RoBERTa outperformed BERT across  all metrics and surpassed GPT-2 in all except AUC. AUC mainly focuses on correctly predicting links, but finding rare true \textcolor{black}{links} is more crucial due to the data imbalance in ILR. RoBERTa \textcolor{black}{may have} made more errors on some false \textcolor{black}{links} (\textcolor{black}{resulting in a lower AUC}), but it ensured higher accuracy on true \textcolor{black}{links} (\textcolor{black}{resulting in a higher Recall}). 

In terms of stability (as shown in Figure~\ref{result RQ2_pci_2}), the standard \textcolor{black}{deviations} for RoBERTa and GPT-2 \textcolor{black}{are} less than \textcolor{black}{one-tenth} of their corresponding means across the measures~\cite{Wang2020AnAH}, indicating strong stability and consistency in their performance across projects. BERT also demonstrated stability in all metrics except Recall and MCC. The mean scores of RoBERTa for the six measures are higher than BERT and GPT-2, with \textcolor{black}{F1-score} at 87.42\%, Precision at 93.94\%, Recall at 82.05\%, MCC at 87.28\%, AUC at 87.28\% and ACC in 95.54\%. From a statistical perspective (as shown in Table~\ref{result RQ2_2}), RoBERTa significantly improves over BERT in Precision, MCC, and ACC. When compared to GPT-2, RoBERTa demonstrates significant enhancements across all metrics except for AUC, with the level of significance being higher than ``RoBERTa vs. BERT". Although the improvements in Precision for both ``RoBERTa vs. GPT-2" and ``RoBERTa vs. BERT" are statistically significant (p-value = 0.0313), further analysis reveals that the effect size is larger for GPT-2 compared to BERT. This indicates a more distinct difference between RoBERTa and GPT-2, showing that the performance of the BERT is better than GPT-2.

Although all three models are based on the transformer architecture, RoBERTa differs from BERT by removing the Next Sentence Prediction (NSP) task and being trained with larger batches, more data, and longer sequences. \textcolor{black}{This}  allows RoBERTa to better capture contextual relationships within the data, leading to its superior performance and stability. In contrast, BERT utilizes MLM and the NSP \textcolor{black}{tasks} during pre-training. Since MPLinker only uses MLM through prompt templates, the NSP component in BERT may introduce noise into downstream ILR task, increasing the sensitivity to different datasets and resulting in slightly reduced performance compared to RoBERTa. GPT-2 is a unidirectional transformer model primarily used for text generation. Unlike mask prediction, its training process focuses on the unidirectional language modeling objective. Therefore, GPT-2 performs stable, but the average performance lags behind RoBERTa and BERT. The above analysis provides a good illustration of the impact of inconsistencies \textcolor{black}{between} PLM and ILR \textcolor{black}{tasks}.

\textbf{\textcolor{black}{In conclusion, among the three PLMs, RoBERTa demonstrates the best overall performance, with notable stability, generalization, and consistency. Although BERT outperforms GPT-2, it shows slightly less stability. For subsequent experiments, MPLinker will therefore use RoBERTa as the default model for artifact representation and mask prediction.}}

\subsection{\textcolor{black}{Ablation Study (The result for RQ3)}}
\label{result 3}

\textbf{Motivation.} The proposed MPLinker method enhances \textcolor{black}{link} diversity and prevents model overfitting through adversarial training. Therefore, evaluating the contribution of adversarial training to MPLinker is essential. RQ3 is proposed to explore  explore MPLinker’s key components. This study compares the performance of MPLinker with and without adversarial training.

\textbf{Approach.} \textcolor{black}{RQ3 conducts  ablation experiments on the adversarial training component of MPLinker.} The evaluation uses the same datasets and metrics, employing the Wilcoxon signed-rank test (Woolson 2007) and Cliff’s Delta effect size test for statistical analysis. To objectively assess the performance of the model, adversarial training typically generates \textcolor{black}{link} only on the training set for training process, aiming to make the model resilient to adversarial attacks. The validation and test sets are used to evaluate the model's performance.
 
 \textbf{Result.}
 Table~\ref{table RQ3} shows the ablation study results, with improvements from adversarial training marked in green. \textcolor{black}{``PT" refers to a model trained using only prompt-tuning, and ``PT+ adv" refers to the same model with adversarial training added.} The findings show that adding adversarial training improved performance across all measures for five open-source projects: OODT, Giraph, Keras, Isis, and Nutch, with the most notable gains seen in the Nutch. \textcolor{black}{For log4net, despite a 12.45\% drop in Precision, Recall increased by 13.79\%, F1-score by 1.73\%. Therefore, sacrificing some Precision to improve  F1-score and Recall is acceptable. The MCC decreased by 2.69\%. This may be due to the smaller size of the log4net compared to others. While adversarial training improved the recall of \textcolor{black}{true links}, over-predicting true links affected the overall balance of the model. However, the performance of MPLinker showed improvements in most metrics on the log4net.}  
 
 MPLinker (adv) shows significant improvements in \textcolor{black}{F1-score}, Recall, ACC, and AUC compared to MPLinker without adversarial training, with large effect sizes across all metrics. Although MPLinker without adversarial training performs best on Precision and MCC in the log4net, MPLinker (adv) achieves the highest scores on other datasets. Because log4net has the smallest dataset size, the intensity of adversarial samples may lead to significant distribution changes, thus challenging the model's generalizability and causing some measures reduction. \textbf{Overall, adversarial training enhances the performance of MPLinker, effectively increasing \textcolor{black}{link} diversity and mitigating model overfitting.}

\begin{table}
\begin{center}
\setlength{\tabcolsep}{1pt}
\scriptsize
\caption{\textcolor{black}{A comparative analysis of Prompt-tuning with adversarial training (PT+Adv) and Prompt-tuning without adversarial training (PT) was conducted using Wilcoxon Signed-Rank tests and Cliff's Delta to evaluate the statistical significance of the results. The $\ast$ indicate significance levels of 0.05.}}
\label{table RQ3}
\begin{tabular}{cccccccc}
\hline
\multirow{2}{*}{\textbf{Project}} &\multirow{2}{*}{\textbf{Model}}  && &\textbf{Metric}&& \\
& & \textbf{F1} & \textbf{Precision} & \textbf{Recall} & \textbf{MCC} & \textbf{AUC} & \textbf{ACC} \\
 \hline
\multirow{2}{*}{\textbf{log4net}} & \textbf{PT}
& 86.79	&96.83&	79.31&	86.79&	86.79	&93.4
 \\
 &\textbf{PT+Adv}&88.52( \textcolor{green}{+1.73})&	84.38(-12.45)&	93.1(\textcolor{green}{+13.79})&	84.1(-2.69)&	93.3(\textcolor{green}{+6.51})&	93.4(\textcolor{green}{+0.00})
\\ \hline
\multirow{2}{*}{\textbf{OODT}} & \textbf{PT} 
& 86.32	&90.99&	82.11	&84.64&	90.48	&96.77
 \\
 &\textbf{PT+Adv}&94.42(\textcolor{green}{+8.1})&	100(\textcolor{green}{+9.01})	&89.43(\textcolor{green}{+7.32})&	93.87(\textcolor{green}{+9.23})	&94.72(\textcolor{green}{+4.24})&	98.69(\textcolor{green}{+1.92})
\\
 \hline
\multirow{2}{*}{\textbf{Giraph}} & \textbf{PT}
& 88.89	&98.55&	80.95	&88.15	&90.4&	97.71
 \\
 &\textbf{PT+Adv}&99.41(\textcolor{green}{+10.52})&	98.82(\textcolor{green}{+0.27})	&100(\textcolor{green}{+19.05})	&99.33(\textcolor{green}{+11.18})&	99.92(\textcolor{green}{+9.52})	&99.87(\textcolor{green}{+2.16})
\\
 \hline
 \multirow{2}{*}{\textbf{Keras}} & \textbf{PT}
& 91.55&	92.86&	90.28&	91.67&	91.67&	91.67
 \\
 &\textbf{PT+Adv}&99.31(\textcolor{green}{+7.76})&	98.63(\textcolor{green}{+5.77})&	100(\textcolor{green}{+9.72})&	98.62(\textcolor{green}{+6.95})&	99.31(\textcolor{green}{+7.64})&	99.31(\textcolor{green}{+7.64})
 \\
 \hline
  \multirow{2}{*}{\textbf{Isis}} & \textbf{PT}
& 88.90 &	93.09 &	85.07& 	92.20 &	87.89 &	97.97 
 \\
 &\textbf{PT+Adv}&99.12(\textcolor{green}{+10.22})&	98.26(\textcolor{green}{+5.17})&	100(\textcolor{green}{+14.93})&	98.99(\textcolor{green}{+6.79})&	99.12(\textcolor{green}{+11.23})&	99.77(\textcolor{green}{+1.80})
 \\
 \hline
   \multirow{2}{*}{\textbf{Nutch}} & \textbf{PT}
&82.08& 	91.30& 	74.56& 	80.21& 	86.74& 	95.73 
 \\
 &\textbf{PT+Adv}&95.83(\textcolor{green}{+13.75})&	98.83(\textcolor{green}{+7.53})&	93(\textcolor{green}{+18.44})&	89.32(\textcolor{green}{+9.11})	&89.91(\textcolor{green}{+3.17})&	97.84(\textcolor{green}{+2.11})

 \\
  \hline
 &\textbf{p-value}&	0.0313*& 	0.4375& 	0.0313*& 	0.0625 	&0.0313*& 	0.0431* 
 \\
    &\textbf{Cliff's Delta}&	0.8333 &	0.6111& 	0.9444& 	0.6111& 	0.8333 	&0.6944 
\\
 \hline
\end{tabular}
\end{center}
\end{table}

\begin{table}
 \caption{The performance comparison of MPLinker and the state-of-the-art ILR methods.}
\scriptsize
\centering
\label{result RQ4_1}
\setlength{\tabcolsep}{2pt}
\begin{tabular}{cccccccc}
\hline
\multirow {2}{*}{\textbf{Project}} &\multirow{2}{*}{\textbf{Approach}}  && &\textbf{Measures}&& \\
& & \textbf{F1} & \textbf{Precision} & \textbf{Recall} & \textbf{MCC} & \textbf{AUC} & \textbf{ACC} \\
 \hline
\multirow{5}{*}{\textbf{log4net}} & \textbf{BTLink}& 77.61 	&89.66 	&68.42& 	68.86& 	87.04 &	85.85 \\
 &\textbf{FRLink}&59.57 &	48.28& 77.78	 &	51.15 	&71.54 	&82.08 	\\
& \textbf{DeepLink}&51.28 &	52.63 &	50.00 &	35.61 &	68.12 	&76.25  \\
 & \textbf{hybrid-linker}&51.72& 	51.72& 	51.72 &	33.54 &	66.77 &	73.58 \\
& \textbf{MPLinker}&88.52 	&84.38& 	93.10& 	84.10 &	93.30& 	93.40 \\
\hline
\multirow {5}{*}{\textbf{OODT}} &  \textbf{BTLink}& 78.51 &	79.83 &	77.24 &	75.54 &	87.24 &94.75  \\
 &\textbf{FRLink}&54.63 	&63.44& 	47.97 &	49.81 &	72.03 &	90.11 \\
& \textbf{DeepLink}&47.02& 	36.95 &	64.66 	&38.91& 	74.29 &	81.47 \\
 & \textbf{hybrid-linker}&59.38 &	82.61 &	46.34 	&58.24 &	72.48 	&92.13 \\ 
& \textbf{MPLinker}&94.42	&100.00 &	89.43	&93.87&	94.72&98.69 \\
 \hline
\multirow {5}{*}{\textbf{Giraph}} & \textbf{BTLink}& 89.44 &93.51 &85.71 &88.26 &92.48 &	97.71  \\
&\textbf{FRLink}&69.74& 	77.94& 	63.10 &	66.79& 	80.41 &	93.80 \\
& \textbf{DeepLink}&41.83& 	29.41 	&72.37 &35.49 &75.11&77.23  \\
& \textbf{hybrid-linker}&4.65 &	100.00 &2.38 &14.55& 51.19 &	88.95 \\
& \textbf{MPLinker}&99.41&	98.82&	100.00	&99.33&	99.92	&99.87\\
 \hline
\multirow {5}{*}{\textbf{Keras}} & \textbf{BTLink}& 92.31 &	92.96 	&91.67 &	84.73 &	92.36 &	92.36 \\
&\textbf{FRLink}&71.11 	&76.19 &	66.67& 	46.20 &	72.92 &	72.92 \\
& \textbf{DeepLink}&54.90& 	50.00 &	60.87& 	4.93& 	52.43& 	52.08  \\
& \textbf{hybrid-linker}&70.67 &	67.95 &	73.61 	&39.02 &69.44 &	69.44 \\
& \textbf{MPLinker}&99.31&	98.63&	100.00&	98.62&	99.31	&99.31\\
\hline
\multirow {5}{*}{\textbf{Nutch}}& \textbf{BTLink}& 80.27&	94.40&	69.82&	78.92&	84.60&95.49\\
&\textbf{FRLink}&67.58&	79.84&	58.58	&64.49&	78.17&	92.62\\
& \textbf{DeepLink}&42.95&	32.88	&61.94&	33.16&	71.14&	77.86 \\
& \textbf{hybrid-linker}&12.83&	66.67	&7.10&	18.88&	53.28	&87.33\\
& \textbf{MPLinker}&95.83&	98.83	&93.00	&89.32&	89.91&	97.84\\
\hline
\multirow {5}{*}{\textbf{Isis}}& \textbf{BTLink}& 78.45&	71.37&	87.08&	76.39	&91.70&	95.43\\
&\textbf{FRLink}&26.91&	22.28&	33.97&	18.24	&60.96	&82.96\\
& \textbf{DeepLink}&29.40&	18.65&	69.32&	25.02	&70.93&	72.27\\
& \textbf{hybrid-linker}&73.21	&88.36&	62.49	&72.24	&80.83	&95.77\\
& \textbf{MPLinker}&99.12 &	98.26 &	100.00 &	98.99 	&99.12& 	99.77 \\
\hline
\end{tabular}
\end{table}

\begin{table}
\scriptsize
\centering
\setlength{\tabcolsep}{1pt}
\caption{\textcolor{black}{Statistical analysis of MPLinker and baselines using wilcoxon Signed-Rank Tests and Cliff’s Delta. The $\ast$ indicate significance levels of 0.05. }}
\label{result RQ4_2}
\begin{tabular}{cccccccc}
\hline
\multirow{2}{*}{\textbf{Approaches}} & && &\textbf{Measures}&& \\
&& \textbf{F1} & \textbf{Precision} & \textbf{Recall} & \textbf{MCC} & \textbf{AUC} & \textbf{ACC} \\
\hline
\multirow{2}{*}{\textbf{MPLinker vs BTLink} } 
&p-value&0.0313* 	&0.0938 &	0.0313* 	&0.0313* &	0.0313* &	0.0313* \\
&Cliff's Delta&0.8889 &	0.7778 	&0.9444 &	0.8889 &	0.8333 &	0.7778 \\
\multirow{2}{*}{\textbf{MPLinker vs FRLink}} 
&p-value&0.0313* &	0.0313* &	0.0313*& 	0.0313* 	&0.0313* &	0.0313* \\
&Cliff's Delta&1.0000& 	1.0000 	&1.0000 &	1.0000& 	1.0000& 	0.9444 \\
\multirow{2}{*}{\textbf{MPLinker vs DeepLink}}
&p-value&0.0313* &	0.0313* 	&0.0313* &	0.0313* &	0.0313*& 	0.0313* \\
&Cliff's Delta&1.0000& 	1.0000 &	1.0000& 	1.0000& 	1.0000 	&1.0000 \\
\multirow{2}{*}{\textbf{MPLinker vs hybrid-linker}}  
&p-value&0.0313* &	0.0625& 	0.0313*& 	0.0313*& 	0.0313* 	&0.0313* \\
&Cliff's Delta&1.0000 &	0.6389& 	1.0000 &	1.0000 &	1.0000 &	0.9444 \\
\hline
\end{tabular}
\end{table}

\subsection{\textcolor{black}{Comparison with State-of-the-Art Methods (The result for RQ4)}}

\textbf{Motivation.} Although MPLinker performed well in the above experiments, its effectiveness cannot be fully confirmed. Therefore, RQ4 further compares MPLinker with the state-of-the-art ILR methods.

\textbf{Approach.} To answer RQ4, selecting suitable baseline methods for comparison is necessary. Thus, this study chose the most effective and widely used methods: FRLink~\cite{sun2017frlink}, DeepLink~\cite{ruan2019deeplink}, hybrid-linker~\cite{mazrae2021automated} and BTLink~\cite{lan2023btlink}. The Wilcoxon signed-rank tests and Cliff’s Delta were used to verify its effectiveness. The experiment uses the same datasets to ensure fairness, maintaining their preprocessing methods and data imbalance handling, although MPLinker does not use any data balance techniques.

\textbf{Result.} Table~\ref{result RQ4_1} shows the comparison results between MPLinker and baselines. Table~\ref{result RQ4_2} presents the hypothesis testing results. \textbf{The results indicate that MPLinker
performs the greatest results when compared to baselines on all measures, achieves an average \textcolor{black}{F1-score} of 96.10\%, Precision of 96.49\%, Recall of 95.92\%, MCC of 94.04\%, AUC of 96.05\%, and ACC of 98.15\% across six projects. The effect size between MPLinker and other ILR methods is large in most steps.}

Compared to BTLink, MPLinker significantly improves \textcolor{black}{F1-score}, Recall, AUC, ACC, and MCC(p-value $<$ 0.05). Meanwhile, the effect sizes in all measures between MPLinker and BTLink are large. For Precision, MPLinker is better than BTLink on all five projects except log4net, and although MPLinker in Precision is down by 5.28\% compared to BTLink, its \textcolor{black}{F1-score} and Recall are increased by 10.91\% and 24.68 \%, respectively. Overall, MPLinker performs better than BTLink. The reason is that BTLink~\cite{lan2023btlink} converts the ILR task into a binary classification task, which differs from Roberta's mask prediction. Compared to DeepLink and FRLink, MPLinker shows significant improvement on all measures, and the effect sizes are large. The result is most likely caused by the fact that the ``word2vec" used by DeepLink~\cite{ruan2019deeplink} is limited in the size of the constructed corpus relative to the PLM, and cannot fully understand the semantic relationships between words. FRLink~\cite{sun2017frlink} uses only ``TF-IDF" for artifact representation, resulting in limited mining of semantic information about the artifacts. 
Compared to the hybrid-linker~\cite{sun2017improving}, MPLinker shows significant improvements across five metrics: \textcolor{black}{F1-score}, Recall, MCC, AUC, and ACC. Moreover, the effect sizes for all these metrics between MPLinker and hybrid-linker are substantial. Although there is no significant difference in Precision between MPLinker and hybrid-linker, hybrid-linker achieves a Precision of 100\% on the Giraph with good results, but the sacrifice of Recall is too high (2.38\%).

\textbf{In conclusion, MPLinker achieves superior performance, with significant improvements across five evaluated measures except Precision. It shows the superiority of PLM over ``TF-IDF" and ``word2vec" in artifact representation, and the consistency of its downstream tasks enhances the mining of the artifact semantics.}

\section{\textcolor{black}{Discussion}}
\label{disscussion}

\textcolor{black}{This section introduces the potential threats from four perspectives: construct validity, internal validity, external validity, and conclusion validity. The aim is to identify the limitations of the proposed method and experiments and to discuss strategies for mitigating these potential threats. Additionally, we provide a usage guide for MPLinker which helps users use the model in practical applications.}

\subsection{Threats to Validity}
\label{Threats to validity}

\textbf{Construct validity:} We designed three  \textcolor{black}{architectures} for different \textcolor{black}{prompt} strategies: Single-template Prompt, Multi-template Prompt, and CLSPrompt. For the single-template prompt, we specifically designed three different single-template contents. However, the design of these templates cannot reflect all possible templates. \textcolor{black}{Despite this limitation, we ensured the rationality of designing the templates by carefully selecting and constructing them based on established methodologies and relevant literature ~\cite{chen2022knowprompt, Zhang2023PromptLF, shin2020autoprompt}.} Future work will focus on the effect of more types of templates on ILR.

\textbf{Internal validity:}
The conclusions derived from different evaluation metrics across various models may vary. Therefore, to ensure consistency in the conclusions as much as possible, six commonly used metrics were employed to evaluate the ILR performance of the models. Recall, Precision and \textcolor{black}{F1-score} are the most commonly measures in ILR~\cite{ruan2019deeplink, xie2019deeplink, le2015rclinker}. Other measures are commonly used in classification models. 

\textbf{External validity:} To ensure the model's consistency across different datasets, this study evaluated experiments using six popular projects. These projects utilize different programming languages and vary in scale and domain. Furthermore, we cannot guarantee that RoBERTa is the optimal choice among all PLMs. To imitate this problem, we refer to previous studies for effective PLM selection~\cite{liu2019roberta, bert, gpt2}, and evaluate the impact of three PLMs on MPLinker through experiments. The evaluated RoBERTa and BERT have demonstrated excellent performance when applied in the ILR domain~\cite{lin2021traceability, lan2023btlink}. At the same time, GPT-2 has also been utilized in areas such as text generation~\cite{qu2020text}, static malware detection~\cite{demirci2022static}, \textcolor{black}{and program repair  ~\cite{lajko2022towards}}.

\textbf{Conclusion validity:} 
RoBERTa in MPLinker shows the best generalization and performance, as validated by RQ1 through the Wilcoxon Signed-Rank Test. \textcolor{black}{Furthermore, RQ2 to RQ4 employ the Wilcoxon Signed-Rank Test and Cliff’s Delta to analyze the results further.} This ensures the reliability and statistical significance of the results. \textcolor{black}{In the ablation study on adversarial training (Section \ref{result 3}), it demonstrated that the perturbations generated by adversarial training produced favorable results across large-scale datasets. However, for smaller datasets, adversarial training tends to impact the increase in recall, showing a certain bias.}

\subsection{\textcolor{black}{Guidelines for Using the Approach}}
\textcolor{black}{\textbf{Design of Prompt Templates.}
We recommend that, when designing prompt templates, users clearly specify task requirements and include keywords related to downstream tasks, such as issues and commits. This helps the model better understand the context, enhancing the relevance and effectiveness of the generated results. Different prompt templates exhibit varying effectiveness across distinct data scenarios, and therefore, a universal template suitable for all datasets does not exist. Accordingly, we recommend adopting the multi-template prompt method proposed in this paper. This approach combines the outputs of various templates, achieving an optimal balance across diverse scenarios. Such a multi-template prompting design allows the model to adjust more flexibly based on specific data characteristics, avoiding the limitations of single-template approaches and enabling the model to maintain high adaptability and generalization across multiple contexts.}

\textcolor{black}{\textbf{Use Cases for Adversarial Training.}
Adversarial training can significantly enhance a model’s robustness to noise and perturbations during the training process. However, for scenarios with relatively small datasets (e.g., where the number of true and false links is fewer than 1000), the noise introduced by adversarial training may cause the model to bias towards recall-focused predictions, potentially impacting its overall accuracy. Therefore, we recommend applying adversarial training on larger datasets to fully leverage its benefits for model stability. This approach helps reduce noise interference during training while enabling the model to perform better in real-world scenarios with richer data support.}

\section{Conclusions}
\label{Conclusion}
Traceability between issues and commits is crucial in software maintenance. However, with the advent of pre-trained models, researchers have started using pre-trained language models to address the \textcolor{black}{problem} of small sizes and semantic gaps between artifacts. Existing methods often involve meticulously designing neural networks for ILR following the PLM, leading to an inconsistency between the fine-tuning of the model and the PLM. This results in the semantics within the PLM not being fully utilized. To mitigate this problem, this research explores the effectiveness of different prompt strategies in ILR and proposes a prompt-tuning-based ILR method with adversarial training. Experiments explore the effectiveness of the Multi-template architecture on six projects, verifying that adversarial training is effective in preventing model overfitting. Specifically, compared with the state-of-the-art ILR methods(BTLink), it outperforms by 13.34\% (96.10\%, 82.77\%) on \textcolor{black}{F1-score}, 9.65\% (96.49\%, 86.96\%) on Precision, 15.93\% (95.92\%, 79.99\%) on Recall, 15.37\% ( 94.04\%, 78.78\%) on MCC, 6.81\% (96.05\%, 89.24\%) on AUC and 4.55\% (98.15\%, 93.60\%) on ACC. These results show that our proposed MPLinker is effective and generalizable in link recovery, highlighting the potential of the prompt-tuning-based method for ILR.

The future work will focus on three directions: (1) In addition to hard prompts, different soft prompt templates will be designed to explore their effectiveness in performing ILR, specifically by dynamically generating different ILR models through PLM. (2) Further research will investigate prompt-tuning methods across various software artifacts. (3) More strategies will be developed to ensure consistency between ILR and PLM.

\section*{Acknowledgements}
This work is supported by the National Natural Science Foundation of China (No. 62102291), and the Opening Foundation of Engineering Research Center of Hubei Province for Clothing Information (No. 2022HBCI02).




 \newpage
\bibliographystyle{elsarticle-num}
\bibliography{cas-refs}





\end{document}